\documentclass[nohyper,notoc]{JHEP}
\usepackage{epsfig}
\def\be{\begin{equation}}
\def\ee{\end{equation}}
                         \def\bearr{\begin{eqnarray}}
                         \def\eearr{\end{eqnarray}}
\def\benum{\begin{enumerate}}
\def\eenum{\end{enumerate}}
\def\bitem{\begin{itemize}}
\def\eitem{\end{itemize}}

\def\lsim{\:\raisebox{-0.5ex}{$\stackrel{\textstyle<}{\sim}$}\:}
\def\gsim{\:\raisebox{-0.5ex}{$\stackrel{\textstyle>}{\sim}$}\:}
\def\go{\rightarrow}
\def\goes{\longrightarrow}

\def\mET{E_T \hspace{-1.1em}/\;\:}

                         \def\G0{\widetilde G}
                         \def\N0{\widetilde \chi^0}

                         \def\Cpm{\widetilde \chi^\pm}

                         \def\sc{\widetilde c}

\def\sl{\widetilde \ell}
\def\se{\widetilde e}

\def\smu{\widetilde \mu}

\title{Linear Collider Signal of a Wino LSP in Anomaly Mediated
Scenarios}
\author{Dilip Kumar Ghosh, Probir Roy, and Sourov Roy \\
                      Department of Theoretical Physics\\
                       Tata Institute of Fundamental Research \\
                        Mumbai 400 005, India \\
\email{{\it E-mail}: dghosh@theory.tifr.res.in, probir@tifr.res.in, 
         sourov@theory.tifr.res.in}}
\abstract{Selectron (smuon) pair-production in a next generation Linear
Collider, yielding a fast electron (muon) trigger, a visible heavily
ionizing track and/or a resolved soft pion impact parameter and overall
$\mET$, is shown to provide a smoking gun signature for Anomaly Mediated
Supersymmetry Breaking models with a neutral Wino as the Lightest
Supersymmetric Particle, nearly mass-degenerate with the lighter chargino.}
\preprint{TIFR/TH/00-18 \\ hep-ph/0004127}

\begin{document}
Understanding how supersymmetry breaks in the real world from a 
deeper, more fundamental, standpoint is a challenge in high energy physics 
today. An interesting recent idea in this direction has been that of 
Anomaly Mediated Supersymmetry Breaking (AMSB) \cite{randall}, based on 
which a whole class of supersymmetric models [\ref{randall} -- \ref{dedes}] 
have emerged. A crucial signal in a high energy Linear Collider, namely 
$e^+ e^- \rightarrow e^\pm (\mu^\pm) + soft ~\pi^\mp + \mET$, to test 
such a scenario, is proposed in this Letter.

AMSB models are strongly motivated by String Theory which is defined in a 
higher dimensional spacetime and is valid at a very high energy scale. It is
quite natural from that point of view to expect a low energy description of 
the physical world in four dimensions to inherit some of the features of the 
higher dimensional theory. This is indeed the case with AMSB scenarios. AMSB 
occurs when in such a higher dimension, one has a supergravity theory defined 
on two separated parallel 3-branes ($3 + 1$ dimensional subspaces) in a way 
that the Standard Model (SM) particles are localized on one of these while 
the supersymmetry breaking sector is localized on the other. There are no 
tree-level couplings between these two branes and thus the supersymmetry 
breaking sector is truly hidden. Gravity propagates in the bulk and the 
breakdown of supersymmetry is communicated from the hidden to the visible 
sector through the loop-induced super-Weyl anomaly. In the absence of 
tree-level interactions between the two 3-branes, this is the dominant 
contribution to the soft supersymmetry breaking parameters determining the 
masses of various superparticles. In the more commonly used Gravity Mediated 
Supersymmetry Breaking scenario \cite{nilles}, supergravity interactions 
directly communicate supersymmery breaking between the hidden and observable 
sectors at the tree level, so that loop-induced contributions from the 
super-Weyl anomaly, though present, are subdominant. A characteristic 
feature of AMSB models is that the stable LSP or {\em Lightest Supersymmetric 
Particle} ($\tilde \chi^0_1$) is almost exclusively a neutral Wino which is 
nearly mass-degenerate with the lighter chargino ($\tilde \chi^\pm_1$), also 
predominantly a Wino. Models with only AMSB have the problem of tachyonic 
sleptons; however, modified versions exist in which the sleptons have 
physical masses. 

Though the quest for supersymmetry has been a major preoccupation of 
collider experimenters and phenomenologists alike, most of the searches 
and simulation studies so far have been based on the assumption of the 
LSP being predominantly a Bino; i.e. the superpartner of the $U(1)_Y$ gauge 
boson. Various produced superparticles are expected to decay 
into the LSP accompanied by other particles of the Standard Model (SM). The 
LSP escapes detection carrying off missing transverse energy or $\mET$ which 
becomes the classic signature. There have, however, been a few papers 
[\ref{chen} -- \ref{paige}] of late which have considered detection 
possibilities for scenarios in which a largely Wino LSP occurs. Our 
investigation belongs to this genre. We consider the pair production of 
left-selectrons (smuons) in $e^+e^-$ interactions, followed by their 
decays\footnote{These decay channels need not make all of the selectron 
(smuon) width. There are regions of parameter space where 
$\tilde \chi^{\pm,0}_2$ can be lighter than the selectron (smuon), but 
decays like $\tilde e (\tilde \mu) \longrightarrow e (\mu) \tilde \chi^0_2$, 
$\nu \tilde \chi^\pm_2$, which open up there, are only a few percent of 
the branching ratio. Even so, we take these into account in our 
calculations.}
$\se (\smu) \go e (\mu) + \N0_1$, $\se (\smu) \go \nu 
+\Cpm_1$; $\Cpm_1$ will further decay into $\N0_1 +\pi^\pm$. Finally, 
there will be a fast $e^\pm (\mu^\pm)$ trigger, a displaced vertex which can 
be inferred from the impact parameter of a visible soft $\pi^\pm$ and/or a 
heavily ionizing track with high momentum (i.e. nearly straight in the 
magnetic field) and large $\mET$. Similar considerations just with 
selectrons can also be made for $e^- e^-$ collision.
 
The chargino and the neutralino masses, in any version of the
Minimal Supersymmetric Standard Model (MSSM) \cite{mssm1}, are controlled
by the following supersymmetry  parameters at the weak scale: the Bino mass 
$M_1$, the Wino mass $M_2$, the Higgsino mass parameter $\mu$ and the ratio 
$\tan\beta $ of the two Higgs VEVs. The situation, with the LSP 
($\N0_1$) being largely the neutral Wino, obtains when one has
\be
|M_2| < |M_1| \ll |\mu|.   
\label{eq:m2m1}
\ee
One should emphasize that, within a MSSM framework, the mass-hierarchy 
(\ref{eq:m2m1}) is very characteristic of AMSB models. For instance, in such 
models, after taking into account next-to-leading order corrections to the 
gaugino mass parameters, one finds \cite{wells} that $M_1:M_2\simeq 2.8:1$ 
as contrasted with $M_1:M_2 \approx  1:2$ in gauge or usual supergravity 
mediated supersymmetry breaking models with gaugino masses unified at the 
grand unifying scale. The next-to-lightest superparticle in the AMSB case 
is the lighter chargino ($ \Cpm_1 $) which is almost exclusively a Wino. 
Then the masses of $\N0_1$ and $\Cpm_1 $ are verly close and the small 
mass-splitting $\Delta M $ has the form \cite{wells}:
\bearr
\Delta M = \frac{ M_W^4 \tan^2\theta_W}{(M_1 - M_2)\mu^2 } \sin^2 2\beta  
\bigg [1+ {\cal O } \left(\frac{M_2}{\mu},\frac{M^2_W}{\mu M_1} 
\right)\bigg ] \nonumber
      \\[1.2ex]
+ \frac {\alpha M_2}{\pi\sin^2\theta_W}\bigg
[f\left(\frac{M_W^2}{M_2^2} \right)- \cos^2\theta_W f\left(\frac{M_Z^2}
{M_2^2}\right)\bigg ], 
\label{eq:delm}
\eearr
with 
\bearr
f(x) &\equiv & -\frac{x}{4}+\frac{x^2}{8}\ln(x) +\frac{1}{2}\left(1+\frac{x}{2}
 \right) \sqrt{4x-x^2}\bigg[\tan^{-1}\left(\frac{2-x}{\sqrt{4x-x^2}}
\right) \nonumber \\ & &\hspace*{2.5in} - \tan^{-1}\left(\frac{x}
 {\sqrt{4x-x^2}} \right)\bigg ].\nonumber
\eearr
The second term in the RHS of Eq.~\ref{eq:delm} is the one-loop 
contribution which is dominated by gauge boson loops. 

The mass-splitting $\Delta M$ of Eq.~\ref{eq:delm} 
has been investigated numerically \cite{feng,wells} in various region of the
parameter space consistent with Eq.~\ref{eq:m2m1}. The general conclusion 
is that
\be
165~{\rm MeV} ~\lsim ~\Delta M ~\lsim ~1~{\rm GeV}, 
\label{eq:rangem1}
\ee
with $\lim_{\mu \go \infty} $ $\Delta M $ being $165$~MeV. On the other hand,
if a radiative electroweak (EW) symmetry breakdown is sought to be implemented 
in the  AMSB scenario, the ratio $ |\mu/M_2| $ has to be \cite{wells} 
approximately between 3 and 6. Given the LEP lower limit \cite{LEP} of 56 GeV 
on the mass of the lighter chargino {\em for nearly degenerate} 
$\Cpm_1, \N0_1$ (in the anomaly-mediated Wino LSP scenario), it then 
follows that the upper limit on $\Delta M$ cannot be much in excess of 
800 ${\rm MeV}$. In that case we can take 

\begin{equation}
165 ~{\rm MeV} ~\lsim ~\Delta M ~\lsim ~800 ~{\rm MeV}.  
\label{eq:rangem2}
\end{equation}

\noindent Eq.~\ref{eq:rangem2} means that the decay $\Cpm_1 \goes \N0_1 
+ \pi^\pm $ is kinematically allowed. The corresponding branching ratio
is found to vary in the range $93\%-96\%$, the balance being largely due 
to the decay modes $\tilde \chi^\pm_1 \rightarrow \tilde \chi^0_1 + e + 
\nu_e$, $\tilde \chi^0_1 + \mu + \nu_\mu$. The resulting soft pion with a 
sub-GeV energy may be detectable, in which case its impact parameter may 
allow one to infer a displaced vertex. On the other hand, the $\Cpm_1$ may 
have a long enough decay length to show a high momentum heavily ionizing 
track which stops in some of the layers in the vertex detector. The 
experimental issues concerning methods of observing this decay have been 
discussed in the third paper of Ref. \cite{chen} and in Refs. \cite{randall, 
wells}. If the decay length\footnote{Here $c\tau = c \hbar p_{\tilde \chi^\pm}
{(M_{\tilde \chi^\pm} \Gamma_{\tilde \chi^\pm})}^{-1}$ with 
$p_{\tilde \chi^\pm}$, $M_{\tilde \chi^\pm}$ and 
$\Gamma_{\tilde \chi^\pm}$ respectively being the momentum, mass and 
width (all in GeV) of the chargino.} 
$c\tau$ of $\Cpm_1$ is greater than 
$3$ cm., it could be observable\footnote{A CCD or APS vertex detector of 
radius 2.5 cm and a beam pipe of radius 2 cm, have been proposed 
\cite{ijmpa} for TESLA.  The chargino track should be identifiable if it 
covers several layers and also ends in the vertex detector.} 
though the $\pi^\pm$ may be 
too soft to be detected. Contrariwise, if $c\tau < 3 $ cm, the track may not 
be observable but the soft charged pion is likely to be visible with its 
impact parameter $b$ resolved. Thus the event, proposed by us, can be 
triggered by the fast charged lepton emanating from the decay of one of the 
sleptons while it can be identified uniquely in terms of the displaced vertex 
determined by the heavily ionizing charged track, which should be nearly 
straight in the magnetic field because of the high momentum, and/or the 
impact parameter $b$ of the soft pion coming from the two-step decay of the 
other slepton. 

In the anomaly mediated case, gaugino masses are proportional to the 
coefficients of the one-loop beta functions of the corresponding gauge 
couplings (generically denoted as $g$), while scalar masses are determined 
in terms of anomalous dimensions and beta functions of both gauge and
Yukawa couplings (generically denoted as $y$). The expressions 
for the anomaly induced contributions to the soft masses are 

\begin{equation}
M_\lambda = {\beta_g \over g} m_{3/2},
\label{eq:gauginom}
\end{equation}

\begin{equation}
m^2_{\tilde f} = -{\frac 1 4} \left(\frac {\partial \gamma} {\partial g}
{\beta_g} + \frac {\partial \gamma} {\partial y} {\beta_y}\right) 
m^2_{3/2},
\label{eq:scalarm}
\end{equation}

\begin{equation}
A_y = {\beta_y \over y} m_{3/2},
\label{eq:ays}
\end{equation}
where gaugino masses are denoted by $M_\lambda$, scalar masses are given
the generic symbol $m_{\tilde f}$, $m_{3/2}$ is the mass of the
gravitino which here is quite heavy ($\sim$ tens of TeV) and $A_y$ are the 
trilinear soft parameters defined with the convention of the third paper of 
Ref. \cite{moroi}. The renormalization group beta and gamma 
functions are defined as
$\gamma(g,y) \equiv {{d {\ln Z}}/{dt}}$,
~${\beta_g}(g,y) \equiv {{dg}/{dt}}$ ~{\rm and} 
~${\beta_y}(g,y) \equiv {{dy}/{dt}}$,
$t$ being the logarithmic scale variable.

The most striking feature of this AMSB scenario is the invariance of the
expressions for soft SUSY breaking mass parameters Eqs.
(\ref{eq:gauginom} -- \ref{eq:ays}) under renormalization 
group (RG) evolution. Thus, these parameters can be evaluated at any scale 
with the appropriate values of the gauge couplings at that particular scale. 
However, the mass squares of the sleptons, calculated in this way, turn
out to be negative. These tachyonic sleptons constitute a major problem 
of this scenario. The most simple and economical way by which these slepton 
mass-squares can be made positive is to add \cite{randall} a common $m^2_0$ 
to all scalars and this is what we consider. However, our signal is also 
present for models \cite{jack,carena} where this positive term is nonuniversal 
and arises from the D-term of a broken $U(1)$ gauge symmetry. Of course, the
addition of any such term destroys the RG invariance of 
Eq.~\ref{eq:scalarm}. Then, in order to get the correct values of the 
mass-squares of the scalars at the EW scale, one must take into 
account the RG evolution of these soft masses from a very high scale. In our 
calculations we have taken this to be the unification scale 
($\approx 1.5 ~to ~2.0 \times 10^{16} ~{\rm GeV}$) where 
all the three gauge couplings meet and the evolution of these couplings 
reproduces the measured values at the EW scale with $\alpha_s \simeq 0.118$. 
The evolution of gauge and Yukawa couplings has been determined by two-loop 
RG equations. The detailed expressions for scalar and gaugino masses as well 
as the trilinear A-parameters are given in Refs. \ref{wells} and \ref{moroi}. 
The Higgsino mass parameter $\mu$ has been computed using complete one-loop 
correction terms of the effective potential at the scale $Q$ in such a way 
that it reproduces the correct pattern of EW symmetry breaking with $Q$ 
chosen to be the geometric mean of the $t$-squark masses 
$\sqrt{m_{{\tilde t}_1} m_{{\tilde t}_2}}$.  The supersymmetric correction to 
the mass of the bottom quark (sizable for large $\tan\beta$) has also been 
computed to one-loop. We have, moreover, accounted for the constraints 
coming from charge and color conservation as well as from the experimental 
lower limits \cite{expt} on various sparticle masses including 
$m_{\tilde \chi^\pm_1} > 56$~GeV \cite{LEP} and also from the requirement of 
the stau not being the LSP. 
 
We have determined the slepton and chargino/neutralino sector of the MSSM 
mass spectrum completely in terms of $m_{3/2}$, $m_0$, $\tan\beta$ (the 
ratio of the two Higgs vacuum expectation values) and the sign of $\mu$. We 
have checked that our results agree with those of previous authors 
\cite{wells, moroi} for 
$\tan\beta = 3$ with $\mu < 0$ and $\mu > 0$ as well as for $\tan\beta = 30$ 
with $\mu < 0$ and $\mu > 0$. The LSP $\tilde \chi^0_1$ and the lighter 
chargino $\tilde \chi^\pm_1$ are found to be very nearly degenerate, as 
suggested by Eq. \ref{eq:delm}. Indeed, we find $\Delta M$ not only to 
obey the inequality (\ref{eq:rangem1}); but also to be a decreasing function 
of $m_{3/2}$, asymptotically reaching the lower bound of (\ref{eq:rangem1}) 
when the latter gets very high. This function is quite insensitive to the 
value of $m_0$. The left and right selectron masses are also found to be 
almost degenerate. The tiny mass-difference between the latter comes mainly 
from one-loop corrections at the electroweak scale since the anomaly induced 
as well as $D$-term contributions are negligible in comparison. This, again, 
is a distinguishing feature of the AMSB scenario which is based on the 
assumption of a universal contribution to the mass-squared for scalars added 
to make the sleptons non-tachyonic. An important point is that the region of 
parameter space where the masses of the selectrons (smuons) do not lie 
between those of $\tilde \chi^\pm_1$ and $\tilde \chi^{\pm,0}_2$, the latter 
being the higher chargino/neutralino,  is not small, though this does not 
affect\footnote{See, footnote 1.} our analysis. The other important aspect 
of the superparticle spectrum in such an AMSB scenario is that the squarks 
are significantly heavier (typically by at least a factor of four) than the 
sleptons, the squark masses being pushed up by the QCD coupling. This means 
that sleptons should be easier to discover in such models. This is why we 
have chosen to study slepton pair-production in a linear collider. Of course, 
one could also directly study the pair-production of charginos $\tilde
\chi^\pm_1$, each decaying into $\tilde \chi^0_1$ and a soft pion.
However, one would then need to have an additional \cite{chen} hard
initial-state-radiated (ISR) photon to act as a trigger. The event rate
there would be significantly less than that of slepton pair-production
on account of the former process being radiative.  

We have calculated the left-selectron (these would be mass eigenstates 
because of negligible left-right mixing) pair production cross section at 
an $e^+ e^-$ CM energy of $1$~TeV for two values of $\tan\beta$, namely,
$3$ and $30$, for $\mu < 0$ and $\mu > 0$. We have then
folded into it the branching fractions for the decays mentioned in the
first paragraph. The selection cuts that have been used on the decay 
products are as follows : (1) the transverse momentum of the lepton 
$p^{\ell}_T > 5$~GeV, (2) the pseudorapidities of the lepton and the 
pion $|\eta| < 2.5$, (3) the electron-pion isolation variable 
$\Delta R = \sqrt{{(\Delta \eta)}^2 + {(\Delta \phi)}^2} > 0.4$, (4) 
the missing transverse energy $\mET > 20$~GeV and $(5)$ $p^\pi_T > 200 $ MeV 
for a detectable soft pion (N.B. the total momentum of the pion is in the 
range of hundreds of MeV). Contour plots in the 
$m_0-m_{3/2}$ plane for various values of cross-sections (in fb) are 
shown in Fig. \ref{contour}.  The shaded regions are excluded by the 
constraints mentioned earlier; in addition, the selectron mass has 
been required not to exceed $500$~GeV which is the kinematic limit 
for observability in a $1$~TeV Linear Collider. 
The allowed region is somewhat smaller for large $\tan\beta$ because of 
stronger left-right mixing in the stau sector. We see that quite interesting 
regions in the $m_0 - m_{3/2}$ plane are covered for cross sections ranging 
from $10 fb$ to $125 fb$. Our signal should thus generate ${\cal O}({10}^{4})$ 
events for an integrated luminosity of $500 ~{(fb)}^{-1}$. These calculations 
have been done with projected TESLA parameters in mind \cite{zerwas}. For a 
scaled down linear collider, e.g. with a CM energy of $500$~GeV and an 
integrated luminosity of $50 ~{(fb)}^{-1}$, we would expect 
${\cal O} ({10}^{3})$ events. 

We have also plotted the decay length $c\tau$ distribution of the chargino
track in Fig.~\ref{decay_l} with the same selection cuts as used in 
Fig.~\ref{contour}; in addition, we have chosen characteristic sample 
values of  $m_0=230$~GeV and $m_{3/2}=43$~TeV, $\tan\beta = 3$ and $\mu < 0$ 
corresponding to $\Delta M= 182.8$~MeV. We observe a plateau in the 
$c\tau$ distribution in the range $8.5$ to $9.9$~cm which can cover several
layers in the vertex detector. Thus there is a reasonable chance of a direct 
observation of the chargino track. The transverse momentum ($p^\pi_T$) and 
the impact parameter $b$ distributions of the soft pion are plotted in 
Figs.~\ref{bpt_pi}a and \ref{bpt_pi}b respectively with the same input 
parameters and selection cuts as for Fig.~\ref{decay_l}. The 
$b$-distribution extends till about $9.9$~cm and peaks at around $b=8.5$~cm.  
It is clear from the $p^\pi_T$ distribution that the minimum $p^\pi_T$ cut 
of $200$~MeV still leaves a substantial part of the allowed phase space for 
study. For such values of $p^\pi_T$, the $3\sigma $ impact parameter 
resolutions are typically \cite{third} ${\cal O}(10^{-1})$~cm. Of course, we 
have chosen a particularly favorable region of the allowed MSSM parameter 
space. The numbers are not always so good in other regions. We have 
nonetheless checked that $b$ is always significantly above the impact 
parameter resolution value. Hence the prospects of resolving the displaced 
vertex by measuring the soft pion impact parameter here are quite high. Let us 
comment finally that, if selectrons are replaced by smuons (with a fast 
muon used as a trigger), event rates are reduced typically by a factor of 
five on account of s-channel suppression.

An alternative MSSM scenario of nearly degenerate $\tilde \chi^0_1$ and 
$\tilde \chi^\pm_1$ (and $\tilde \chi^0_2$ as well) can arise \cite{chen}
when $|\mu| \ll |M_{1,2}|$. In such a case a mass-difference $\Delta M 
(\tilde \chi^\pm_1 - \tilde \chi^0_1) \lsim 1$~GeV can be obtained with 
$m_{\tilde \chi^\pm_1} > 51$~GeV \cite{maltoni} by setting \cite{chen} 
$|M_{1,2}| \gsim 5$~TeV and $|\mu| \gsim {M_Z}/2$. Though this is a rather 
unnatural scenario and quite difficult to obtain in a phenomenologically 
viable model, we can ask whether our signal can be mimicked here. The answer 
is no. The two-body decays of selectrons, relevant for us, are highly 
suppressed in this other scenario on account of the factor ${m_e}/{M_W}$ 
in the concerned couplings. The latter arises because $\tilde \chi^\pm_1$, 
$\tilde \chi^0_{1,2}$ are all almost exclusively higgsinos here. So 
selectrons primarily have three-body decays $\tilde e \rightarrow 
\nu_e W \tilde \chi^0_{1,2}$, $e Z \tilde \chi^0_{1,2}$ mediated by virtual 
heavier charginos/neutralinos $(\tilde \chi^\pm_2/\tilde \chi^0_2)$, which are 
gauginos, with finals states dominated by jets. One can easily estimate the
ratio of the partial widths of left selectron decays into two-body and 
three-body channels to be  ${\cal O}(10^{-4})$ in this scenario 
demonstrating that the desired two-body decays would be unobservable.      
Therefore, unlike the soft pion plus hard ISR photon signal studied in 
Ref. \cite{chen}, our final state of a fast electron (muon) and a soft 
pion distinguishes AMSB models from the light higgsino scenario. We would 
like to highlight this new result which has emerged from the present work. 

Let us also discuss the question of background to our signal. The signal 
can be classified into two categories. There is one in which 
we see a heavily ionizing nearly straight charged track ending with a 
soft pion with large impact parameter and $\mET$, the signal being 
triggered with a fast electron or a muon. In the other case, while the 
other aspects remain the same, one may not see the heavily ionizing 
charged track but the impact parameter of the soft pion can be resolved 
and measured to be large. In the first case the heavily ionizing charged 
track is due to the passage of a massive chargino with a very large momentum. 
Due to this reason the charged tarck will be nearly straight in the presence 
of the magnetic field. One cannot imagine a similar situation in the SM 
with such a nearly straight heavily ionizing charged track due to a very 
massive particle. An ionized charged track can possibly arise from the 
flight of a low energy charged pion, kaon or proton but it will curl 
significantly in the magnetic field. Another distinguishing feature of 
the charged track in our signal is that it will be terminated after a 
few layers in the vertex detector and there will be a soft pion at the end. 
In the second case, where the ionizing track is unseen, possible SM 
backgrounds can come from the following processes: $e^+ + e^- \rightarrow 
\tau^+ + \tau^-$ and $e^+ + e^- \rightarrow W^+ + W^-$. In the case of 
$e^+ + e^- \rightarrow \tau^+ + \tau^-$, one $\tau$ can have the three 
body decay $\tau \rightarrow e \nu_e \nu_\tau$ or $\mu \nu_\mu \nu_\tau$ 
and the other $\tau$ can go via the two body channel $\tau \rightarrow 
\pi + \nu_\tau$. Thus we can have a final state of the type $e (\mu) 
+ \pi + \mET$. Since we are considering an $(e^+ e^-)$ CM energy of 1 TeV, 
and the pion comes from a sequence of two-body production and decay, it 
will have a fixed high momentum much in excess of 1 GeV. This will clearly 
separate this type of background from our signal since in our case the 
resulting pion is very soft with a momentum in the range of hundreds of MeV. 
In the case of $e^+ + e^- \rightarrow W^+ + W^-$ a similar argument follows. 
Here one $W$ can go to $e (\mu) + \nu_e (\nu_\mu)$ and the other one can go 
to $\tau + \nu_\tau$. The $\tau$ can subsequently go to one $\pi$ and a 
$\nu_\tau$, thereby producing the final state $e (\mu) + \pi + \mET$. As we 
have discussed just now, the resulting pion will have a very large 
momentum and again one can clearly separate the background from the signal.          

In conclusion, we claim to have pinpointed a fast electron (muon) trigger,
overall $\mET > 20$ GeV and a displaced vertex emitting a soft pion in the 
final state configuration as a distinct and unique linear collider signal 
of the AMSB scenario with a Wino LSP. A more detailed discussion of this 
as well as other linear collider signals of AMSB models will be given 
elsewhere.
   
This work came out of a study-project on {\it Anomaly Mediated 
Supersymmetry Breaking} at the Workshop on High Energy Physics Phenomenology 
{\it WHEPP-6} ~(Chennai, January, 2000). We thank the organizers as well
as the other members of the project namely D.~Choudhury, S.~King, A.~Kundu, 
B.~Mukhopadhyaya, S.~Raychaudhuri and K.~Sridhar for many fruitful 
discussions. We also thank U.~Chattopadhyay for the use of his codes and 
M.~Maity and N.~K.~Mondal for discussions of experimental issues.
\vspace*{0.3in}

\def\pr#1 #2 #3 { {\it Phys. Rev. }{\bf #1} (#2) #3}
\def\prd#1 #2 #3{ {\it Phys. Rev. }{\bf D #1} (#2) #3}
\def\prl#1 #2 #3{ {\it Phys. Rev. Lett. }{\bf #1} (#2) #3}
\def\plb#1 #2 #3{ {\it Phys. Lett. }{\bf B #1} (#2) #3}
\def\npb#1 #2 #3{ {\it Nucl. Phys. }{\bf B #1} (#2) #3}
\def\prep#1 #2 #3{ {\it Phys. Rep. }{\bf #1} (#2) #3}
\def\zpc#1 #2 #3{ {\it Z. Physik. }{\bf C #1} (#2) #3}
\def\epjc#1 #2 #3{ {\it Eur. Phys. J. }{\bf C #1} (#2) #3}
\def\mpla#1 #2 #3{ {\it Mod. Phys. Lett. }{\bf A #1} (#2) #3}
\def\ijmpa#1 #2 #3{ {\it Int. J. Mod. Phys. }{\bf A #1} (#2) #3}
\def\ptp#1 #2 #3{ {\it Prog. Theor. Phys. }{\bf #1} (#2) #3}
\def\jhep#1 #2 #3{ {\it JHEP }{\bf #1} (#2) #3}

\newpage

\FIGURE[htb]{\epsfig{file=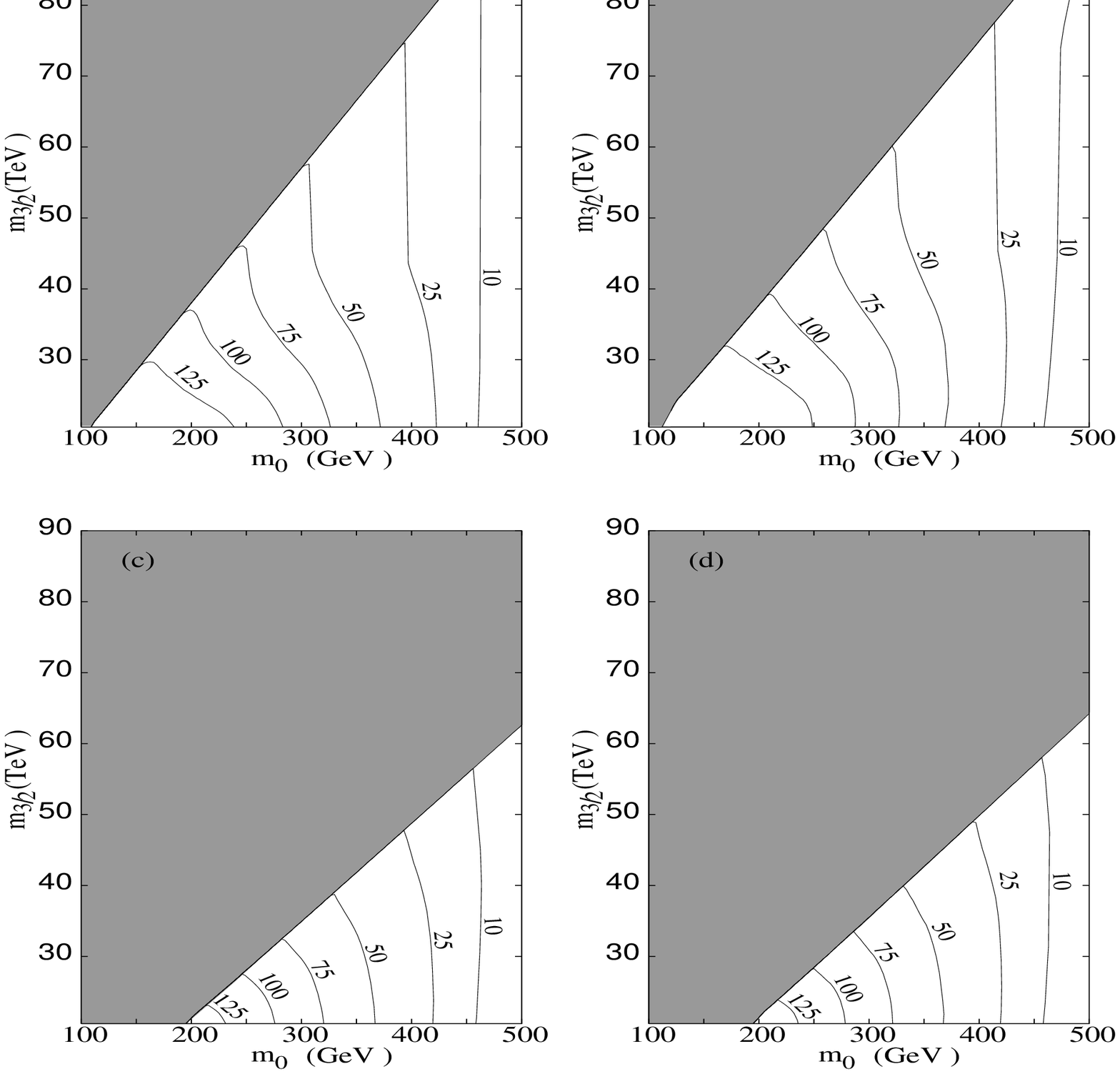,width=10cm}
\vspace*{-1.5in}
\caption{Constant cross section (in {\em fb}) contour plots in the
$m_0-m_{3/2}$ plane for ($a$) $\mu < 0$, $\tan\beta = 3$, ($b$)
$\mu > 0$, $\tan\beta = 3$, ($c$) $\mu < 0$, $\tan\beta = 30$ and
($d$) $\mu > 0$, $\tan\beta = 30$.}
\label{contour}}

\FIGURE[htb]{\epsfig{file=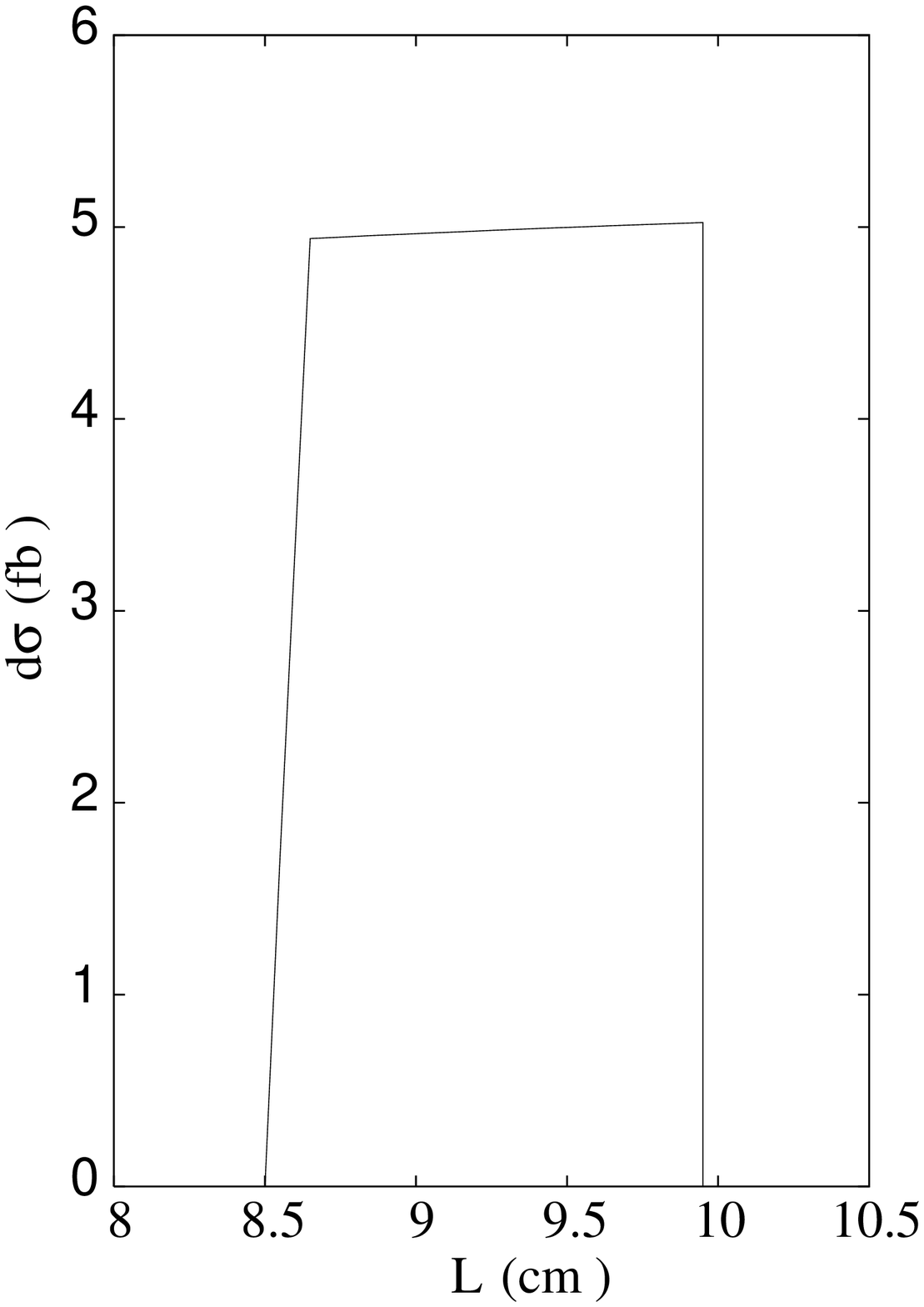,width=10cm}
\vspace*{0.001in}
\caption{Chargino decay length distribution for the following set of
input parameters:~ $m_{3/2}=43$~TeV, $m_0=230$~GeV, $\tan\beta=3$, and
$\mu < 0$.}
\label{decay_l}}

\FIGURE[htb]{\epsfig{file= 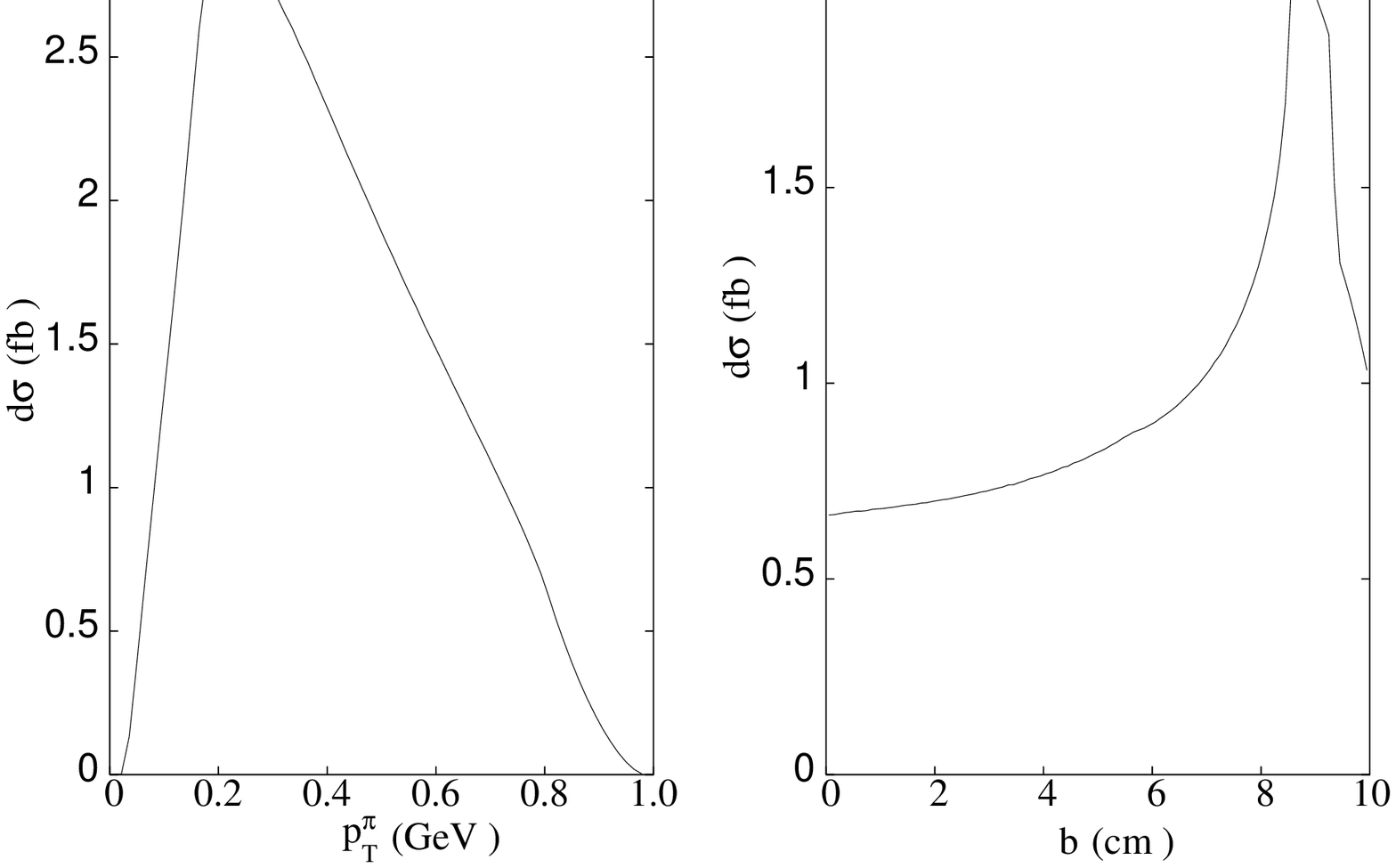,width=10cm}
\vspace*{-2.8in}
\caption{(a) Pion transverse momentum distribution and (b) the impact
parameter distribution for the same set of input parameters as
in Fig.\ref{decay_l}.}
\label{bpt_pi}}

\end{document}